\documentclass[
    aps,
    prl,
    reprint,
    superscriptaddress,
    amsmath,
    amssymb,
    floatfix
  ]{revtex4-2}
\usepackage{amsfonts}
\usepackage{mathrsfs}
\usepackage{amsmath}
\usepackage{color}
\usepackage{natbib}
\usepackage{graphicx}
\usepackage{bm}

\usepackage{xspace}
\usepackage{epstopdf}
\usepackage{dcolumn}
\usepackage{longtable}
\usepackage{array}
\usepackage{makecell}
\usepackage{tabulary}
\usepackage{multirow}
\newcolumntype{C}[1]{>{\centering\arraybackslash}p{#1}}
\newcolumntype{L}[1]{>{\flushleft\arraybackslash}p{#1}}

\usepackage{epsfig}
\usepackage[normalem]{ulem}
\usepackage{braket}
\usepackage{bbold}

\usepackage[colorlinks=true, letterpaper=true, pdfstartview=FitV, linkcolor=blue, citecolor=blue, urlcolor=blue]{hyperref}

\DeclareMathAlphabet{\mathpzc}{OT1}{pzc}{m}{it}

\makeatletter

\newcommand{\Rmnum}[1]{\expandafter\@slowromancap\romannumeral #1@}

\makeatother
\makeatletter
\renewcommand{\paragraph}{%
  \@startsection{paragraph}{4}{\parindent}%
    {\z@}{\z@}%
    {\normalfont\normalsize\itshape}%
}
\makeatother

\begin{document}

\title{Hidden Spin-Valley Hall Effect and Gate-Switchable Electric Hall Effect in 2D Altermagnets}

\author{Jiajun Lu}
\affiliation{Key Lab of advanced optoelectronic quantum architecture and measurement (MOE), Beijing Key Laboratory of Quantum Matter State Control and Ultra-Precision Measurement Technology, and School of Physics, Beijing Institute of Technology, Beijing 100081, China}

\author{Mu Tian}
\affiliation{Key Lab of advanced optoelectronic quantum architecture and measurement (MOE), Beijing Key Laboratory of Quantum Matter State Control and Ultra-Precision Measurement Technology, and School of Physics, Beijing Institute of Technology, Beijing 100081, China}

\author{Chaoxi Cui}
\affiliation{Key Lab of advanced optoelectronic quantum architecture and measurement (MOE), Beijing Key Laboratory of Quantum Matter State Control and Ultra-Precision Measurement Technology, and School of Physics, Beijing Institute of Technology, Beijing 100081, China}

\author{Zhi-Ming Yu}
\affiliation{Key Lab of advanced optoelectronic quantum architecture and measurement (MOE), Beijing Key Laboratory of Quantum Matter State Control and Ultra-Precision Measurement Technology, and School of Physics, Beijing Institute of Technology, Beijing 100081, China}
\affiliation{International Center for Quantum Materials, Beijing Institute of Technology, Zhuhai, 519000, China}

\author{Yugui Yao}
\affiliation{Key Lab of advanced optoelectronic quantum architecture and measurement (MOE), Beijing Key Laboratory of Quantum Matter State Control and Ultra-Precision Measurement Technology, and School of Physics, Beijing Institute of Technology, Beijing 100081, China}
\affiliation{International Center for Quantum Materials, Beijing Institute of Technology, Zhuhai, 519000, China}
\date{\today}

\author{Run-Wu Zhang}
\email{zhangrunwu@bit.edu.cn}
\affiliation{Key Lab of advanced optoelectronic quantum architecture and measurement (MOE), Beijing Key Laboratory of Quantum Matter State Control and Ultra-Precision Measurement Technology, and School of Physics, Beijing Institute of Technology, Beijing 100081, China}
\affiliation{International Center for Quantum Materials, Beijing Institute of Technology, Zhuhai, 519000, China}

\begin{abstract}

Hidden beneath the vanishing charge Hall response of altermagnets are spin-up and spin-down Hall currents of opposite sign, whose cancellation is enforced by symmetry. When these spin states are locked to distinct valleys, the hidden currents constitute a spin-valley Hall effect.
Here we show that in 2D altermagnets with spin-valley-layer coupling (SVLC), a  gate field unveils this hidden spin-valley Hall response.~By lifting the symmetry-enforced compensation between valleys with opposite layer polarizations, the field converts the hidden response into a net charge Hall signal, yielding a gate-switchable electric Hall effect (EHE) enabled by SVLC. 
Using collinear spin layer group (cSLG) theory, we classify all 158 altermagnetic cSLGs, identify 32 that support SVLC, and predict monolayer Mg$_2$Os$_2$(SO$_5$)$_2$ and the Ca(CoN)$_2$-family as candidate materials. First-principles calculations for Mg$_2$Os$_2$(SO$_5$)$_2$ reveal reversible valley splitting, a gate-switchable Hall response, and a Kerr rotation exceeding 1 mrad whose sign reverses together with the EHE signal. These results establish a symmetry-based route to all-electrical valley-spin control and identify altermagnets as promising platforms for high-performance valley transport and electrically switchable magneto-optics.
\end{abstract}
\maketitle

\begingroup
\setlength{\parskip}{0pt}
\paragraph{{\color{blue}Introduction—}}%
Hidden physics~\cite{hidden1,hidden1-1,hidden1-2,hidden1-3,hidden1-4,hidden2} has emerged as a unifying concept in quantum materials: an effect forbidden by the global symmetry can survive in local sectors, with the global response vanishing only through mutual compensation.
Altermagnets~\cite{PRX1,PRX2,AM2,AM-o-1,zhoutong1,zhoutong2,AM1,AM3,AM-o-2,AM-o-3,AM-o-4,AM-o-5,AM-o-6,AM-o-7,zhangjunteng1,zhj1} provide a magnetic realization of this paradigm. They are collinear magnets that combine the compensated magnetic order of antiferromagnets with the spin-split electronic structure of ferromagnets.
Their antiparallel spin sublattices are connected by crystal rotations or mirror operations, rather than by translation or inversion, giving rise to nonrelativistic spin splitting and crystal-symmetry-governed spin-momentum locking~\cite{LJW2,LJW3,wr2,zjt2} while maintaining zero net magnetization. 
This combination suppresses stray fields while retaining spin-polarized electronic responses, making altermagnets promising for scalable and efficient spintronic applications~\cite{xuyin1,xuyin2,liping1,liping2,guosandong1,wxg2,wr1,wxt1}. 
However, the same symmetries often enforce cancellation between the Hall currents carried by opposite-spin states, causing the net charge Hall response to vanish. This vanishing does not indicate an absence of Hall physics but instead reflects a hidden Hall response.

A natural way to distinguish these compensating Hall channels is through the valley degree of freedom.
Valleys are inequivalent energy extrema in momentum space ~\cite{valley1,valley2,VLC1,VLC2,VLC4}. When different valleys carry opposite Berry curvatures, their carriers acquire opposite transverse velocities under an in-plane electric field, producing a valley Hall response. 
When the valley index is locked to spin, these counterpropagating valley currents also carry opposite spin polarizations~\cite{SVHE1,SVHE2,SVHE3}, constituting a spin-valley Hall response that arises naturally in altermagnets with symmetry-related opposite-spin valleys [see Fig.~\ref{fig:one}{\color{blue} (a)}].
This response is hidden in the charge channel: the two valleys contribute opposite Hall signals and cancel exactly, so it cannot be converted into a measurable charge Hall signal unless the two valleys are selectively unbalanced.

Several approaches can lift the valley degeneracy~\cite{VLC4,valley1,valley2,VVV1,VVV2,VVV3}, but electrical control is particularly attractive for practical devices~\cite{EC1,EC2,EC3,EC4,EC5,wxg1,zjt1,sdf2}. Although a gate field does not directly couple to the valley index, the layer degree of freedom provides an indirect route: opposite layer polarizations shift the two valleys in opposite directions and produce reversible valley splitting~\cite{VLC_Yu2020}.
In an altermagnet, combining valley-layer coupling with spin-valley locking associates each valley with both a definite spin character and an opposite layer polarization, giving rise to spin-valley-layer coupling (SVLC)~\cite{SLC_Zhang2024}.
SVLC therefore allows a gate field to lift the valley degeneracy, make one spin-polarized valley dominate transport, and convert the hidden spin-valley Hall response into an electrically detectable charge Hall response, as shown in Figs.~\ref{fig:one}{\color{blue}(b)} and~\ref{fig:one}{\color{blue}(c)}.
This provides a concrete route to an electric Hall effect (EHE) driven by hidden spin-valley physics.
\begin{figure}[htb]
\centering
	\includegraphics[width=\columnwidth]{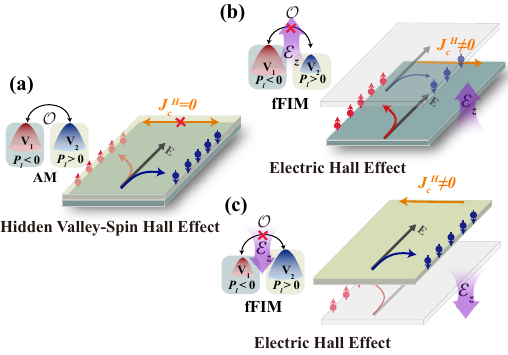}
	\caption{
		 Schematic of the gate-controlled conversion from a hidden spin-valley Hall effect to an electric Hall effect (EHE). (a) Altermagnet (AM) exhibiting a hidden spin-valley Hall effect. The two valleys $V_1$ and $V_2$, which are related by symmetry $\mathcal{O}$, generate opposite transverse valley currents, while the net charge Hall response cancels($J_c^{H}=0$). Fully compensated ferrimagnet (fFIM) exhibiting an EHE under (b) $\mathcal{E}_z>0$, and (c)  $\mathcal{E}_z<0$. The gate field lifts the valley degeneracy: a positive field makes $V_1$ the dominant conducting valley, whereas reversing the field selects $V_2$ and reverses the anomalous Hall response. Here, the sign of $P_l$ indicates the layer polarization: $P_l>0$ ($P_l<0$)  corresponds to a state  predominantly localized on the top (bottom) layer. $J_c^{H}$ is the total charge Hall current, defined as the sum of spin-up and spin-down currents, $J_c^{H}=J^H_{\uparrow}+J^{H}{\downarrow}$.
	}
	\label{fig:one}
\end{figure}

Here, we show that SVLC enables a gate field to convert the hidden spin-valley Hall response of a 2D altermagnet into a gate-switchable EHE~\cite{EHE}.
By lifting the degeneracy of opposite-spin valleys through their opposite layer polarizations, the gate field drives the system into a fully compensated ferrimagnet (fFIM)~\cite{fFIM1,fFIM2,fFIM3} while preserving zero net magnetization. Within collinear spin layer group (cSLG) theory, we derive the symmetry criteria for this mechanism and classify all 158 altermagnetic cSLGs, identifying 32 that support SVLC and predicting monolayer Mg$_2$Os$_2$(SO$_5$)$_2$ and the Ca(CoN)$_2$ family as candidate materials. 
First-principles calculations for Mg$_2$Os$_2$(SO$_5$)$_2$ confirm a zero-field spin-valley Hall response with vanishing charge Hall conductivity, reversible valley splitting, and a Hall response whose sign switches with gate polarity. We further predict a Kerr rotation exceeding 1 mrad that reverses together with the EHE signal. These results establish altermagnets with SVLC as a concrete platform for all-electrical valley-spin manipulation, gate-switchable Hall transport, and electrically reversible Kerr signals.

\endgroup
\paragraph{{\color{blue} Symmetry criteria and cSLG classification—}}%
The hidden spin-valley Hall response is intrinsic to altermagnets in which opposite-spin valleys are related by a spin-group operation. At zero external field, the two valleys carry opposite Berry curvatures and opposite spin polarizations, so their Hall currents compensate.

We define the total charge Hall current as $J_c^{H}\equiv J_{\uparrow}^{H}+J_{\downarrow}^{H}$, where $J_{\uparrow}^{H}$ and $J_{\downarrow}^{H}$ denote the Hall charge-current contributions from the spin-up and spin-down channels, respectively. The symmetry enforces $J_{\uparrow}^{H}=-J_{\downarrow}^{H}$ at $\mathcal{E}_z=0$, yielding $J_c^{H}=0$ even though the two spin-resolved Hall currents are individually finite. The Hall response is therefore hidden in the total charge channel rather than absent.

To make this hidden response experimentally accessible, the valley degeneracy must be lifted. This requires SVLC, which links each valley to a definite spin polarization and an opposite layer polarization. Under these conditions, a perpendicular gate field can break the spin-group operation relating the two valleys, lift their degeneracy through the opposite layer polarizations, and convert the hidden spin-valley Hall response into a gate-switchable EHE. This yields a finite charge Hall current, $J_c^{H}\neq0$.

We first outline the  relevant  symmetry prerequisites:
(i) the transport valleys must be related by a spin-group operation and carry opposite spin polarizations;
(ii) the spatial component of this operation must be broken by an applied gate field;
(iii) the same valleys must exhibit finite and opposite layer polarizations.
When all three conditions are satisfied, the gate field lifts the degeneracy of the opposite-spin valleys via their layer polarization, driving the altermagnet into an fFIM while preserving zero net magnetic moment.

We formalize these criteria using the spin-group operation $\hat{S}=\{ g_s \parallel g_l \}$, where the left and right entries act on spin and real space, respectively. For the valley pair $V_1$ and $V_2$, $g_l V_1=V_2$ and $\hat{S}\ket{\psi(V_1)}\sim\ket{\psi(V_2)}$. The spin-space part $g_s=-1$ reverses the spin expectation value, while the real-space part $g_l=\mathcal{O}$ connects the two valley momenta. Hence, as long as $\hat{S}$ is preserved, the paired valleys carry opposite spin polarizations~\cite{VLC_Yu2020,SLC_Zhang2024}.

For a Bloch state $\ket{\psi(\boldsymbol{k})}$, we define the spin and layer polarizations as $P_{s(l)} = \bra{\psi(\boldsymbol{k})} \hat{P}_{s(l)} \ket{\psi(\boldsymbol{k})}$, with $\hat{P}_s \equiv \hat{s}_z$ and $\hat{P}_l \equiv \hat{z}$. The origin of $z$ is taken at the monolayer midplane, and the ``top'' and ``bottom'' layers denote two effective sublayers, each of which may contain several atomic planes. Thus, $P_s>0$ ($P_s<0$) indicates dominant spin-up (spin-down) weight, while $P_l>0$ ($P_l<0$) indicates dominant top-sublayer (bottom-sublayer) weight. The layer polarization is a property of the Bloch wavefunction in real space and should not be inferred simply from the magnetic sublattice configuration. SVLC further requires valley-contrasted layer polarization, namely $P_s(V_1)=-P_s(V_2)$ and $P_l(V_1)=-P_l(V_2)$.

Since $H_E=e\mathcal{E}_z z$ is spin independent, $g_s$ acts trivially on it; only the real-space part $g_l$ matters. With the external field direction held fixed, we have $\hat{S}H_E\hat{S}^{-1}=e\mathcal{E}_z g_l(z)$. Thus $\hat{S}$ is broken whenever $g_l(z)\neq z$. In the layer-exchange case relevant here, $g_l$ interchanges the two valley states and reverses their layer polarizations, with the ideal limit $g_l(z)=-z$. The electric field then induces valley-dependent first-order energy shifts
$
\Delta E_i=e\mathcal{E}_z\langle z\rangle_{V_i},
$
such that opposite layer polarizations produce a finite valley splitting
$
\Delta E_v=e\mathcal{E}_z(\langle z\rangle_{V_1}-\langle z\rangle_{V_2}),
$
whose sign reverses with $\mathcal{E}_z$. Thus the gate field does not couple directly to the valley label; rather, it breaks the spin-group operation that relates the opposite-spin valleys and exploits their opposite layer polarizations to lift the valley degeneracy. This symmetry reduction converts the parent altermagnet into an fFIM, as shown in Figs.~\ref{fig:one}{\color{blue}(b)} and~\ref{fig:one}{\color{blue}(c)}, and enables a layer-selective gate potential to generate valley splitting that drives spin-valley-selective anomalous transport.

To screen this symmetry relation at the 2D crystallographic level, one must specify which operations are allowed in the 2D magnetic crystal. This is precisely the role of cSLGs, the 2D spin-group extension of ordinary layer groups, adapted to the inequivalence between the in-plane and out-of-plane directions. For clarity and notational convenience~\cite{zhangzeying1,tianmu1,SSG1,SSG2,SSG3,SSG4}, we label a cSLG as $(L.L_{0}.L_{1}.i_{k}.m)$, which compactly encodes its symmetry structure. Detailed definitions of this notation are provided in the {\color{blue} Supplemental Material (SM)}~\cite{SM}.

\begin{figure}[htb]
\centering
\includegraphics[width=\columnwidth]{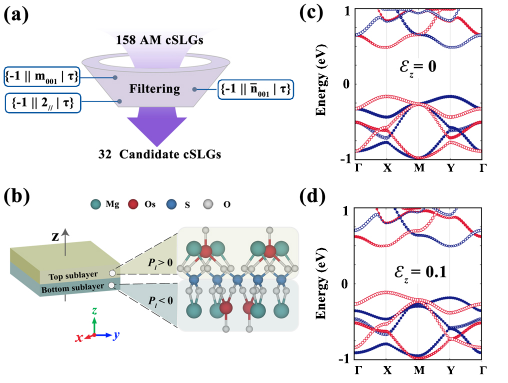}
\caption{(a) Symmetry-screening flow from 158 altermagnetic cSLGs to 32 candidate cSLGs satisfying the SVLC-enabled EHE criteria. (b) Schematic of a 2D material system composed of two effective sublayers, each containing multiple atomic planes. The inset shows a side view of a monolayer Mg$_2$Os$_2$(SO$_5$)$_2$, in which positive (negative) $P_l$ indicates the top (bottom) sublayer. Electronic band structures of monolayer Mg$_2$Os$_2$(SO$_5$)$_2$ at (c) $\mathcal{E}_z=0$ and (d) $\mathcal{E}_z=0.1\,\text{eV/\AA}$, respectively. Blue and red denote spin-down and spin-up states, respectively, while open and filled symbols denote $P_l<0$ and $P_l>0$, respectively. Both band structures are calculated in the absence of spin-orbit coupling (SOC). The corresponding top-view structure, Brillouin zone, and negative-field band structure are provided in the {\color{blue} Supplemental Material}~\cite{SM}.}
\label{fig:two}
\end{figure}

Applying these criteria to the 158 altermagnetic cSLGs reduces the list to 32 cSLGs that satisfy the symmetry requirements for SVLC-enabled EHE [Fig.~\ref{fig:two}{\color{blue}(a)}]. Table~{\color{blue} S2} groups these qualified cSLGs by 2D Bravais lattice type (see {\color{blue}SM}~\cite{SM} for details). This classification offers a symmetry-guided route to identify candidate materials prior to material-specific first-principles calculations.

\paragraph{{\color{blue} Material realization—}}%
We next validate the symmetry principle in monolayer Mg$_2$Os$_2$(SO$_5$)$_2$, which belongs to the qualified cSLG $L.59.23.1.1$. The optimized structure adopts a square lattice with $a=b=5.56$ \r{A}. Along the \textit{z} direction, the monolayer forms a centrosymmetric seven-layer atomic stacking: a central S layer sandwiched by bridging oxygen layers, nearly coplanar Mg-Os metallic layers, and terminal oxygen layers [Fig.~\ref{fig:two}{\color{blue}(b)}]; the top-view crystal structure and corresponding first Brillouin zone are provided in {\color{blue} Fig. S1 of SM}~\cite{SM}.

The electronic structure confirms the SVLC condition required by the screening rule. In the absence of spin-orbit coupling (SOC), both valence and conduction bands exhibit valleys at the $X$ and $Y$ points [Fig.~\ref{fig:two}{\color{blue}(c)}], with predominantly spin-up and spin-down characters, respectively. For the valence band, the $X$ and $Y$ states are mainly localized on the bottom and top sublayers, respectively (Figs.~\ref{fig:two}{\color{blue} (c)} and~\ref{fig:two}{\color{blue} (d)}), establishing valley-contrasted spin and layer polarizations. The symmetry-related valleys have compensating Hall contributions, leaving a hidden spin-valley Hall response with $J_c^{H}=0$. These valley characteristics remain essentially unchanged upon inclusion of SOC ({\color{blue} Fig.~S2}~\cite{SM}).

The gate-field response follows directly from this spin-valley-layer locking. Under an experimentally accessible field $\mathcal{E}_z=0.1$ eV/\r{A}~\cite{E_0.1}, the two valence valleys shift in opposite directions [Fig.~\ref{fig:two}{\color{blue}(d)}]: the $X$ valley moves upward and the $Y$ valley moves downward. Reversing the field to $\mathcal{E}_z=-0.1$ eV/\r{A} reverses this ordering, as expected for a layer-polarization-controlled valley splitting (see Fig.~{\color{blue} S1(c)}~\cite{SM}). In contrast, the conduction valleys exhibit much weaker splitting, as their layer polarizations lack strong valley contrast. This contrast underscores why valley-contrasted layer polarization, rather than mere layer localization, is essential for efficient gate control. The calculated field dependence of the valence-valley splitting further confirms continuous and reversible electric control of valley polarization (Fig.~{\color{blue} S3}~\cite{SM}).

\paragraph{{\color{blue} Gate-switchable electric Hall effect—}}%
\begin{figure}[htb]
\centering
\includegraphics[width=\columnwidth,trim=0 0.4cm 0 0cm, clip]{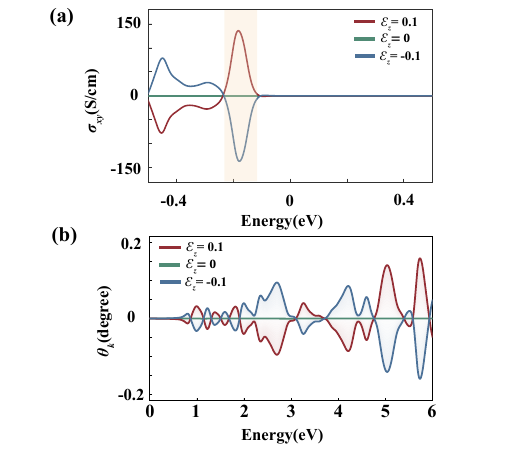}
\caption{ (a) Calculated anomalous Hall conductivity $\sigma_{xy}$ of monolayer Mg$_2$Os$_2$(SO$_5$)$_2$ under gate fields of  $\mathcal{E}_z=0.1$ (red),  $\mathcal{E}_z=0$ (green), and  $\mathcal{E}_z=-0.1$ (blue) eV/\AA, respectively. The yellow shaded regions mark the energy window between the two valence-band extrema. The inset schematics show the relative positions of the Fermi level $E_f$ and the two inequivalent valleys. (b) Calculated Kerr rotation angle $\theta_K$ as functions of energy for monolayer Mg$_2$Os$_2$(SO$_5$)$_2$ under perpendicular electric fields $\mathcal{E}_z=0.1$ (red), $\mathcal{E}_z=0$ (green), and $\mathcal{E}_z=-0.1$ (blue) eV/\AA, respectively.}
\label{fig:three}
\end{figure}


With the gate-controlled valley splitting established, we proceed to investigate the transverse transport response. The Berry curvature $\Omega(\boldsymbol{k})$ generates an anomalous transverse velocity under an in-plane electric field, $v\sim \boldsymbol{E}\times \Omega(\boldsymbol{k})$~\cite{valley2,fcsd2}. We therefore compare the Berry curvature and anomalous Hall conductivity (AHC) under three conditions: $\mathcal{E}_z=0$, $\mathcal{E}_z>0$, and $\mathcal{E}_z<0$.
As for $\mathcal{E}_z=0$ case, the altermagnetic symmetry still relates the $X$ and $Y$ valley states with opposite spin polarizations. The Berry curvature near the $X$ and $Y$ valleys thus has opposite sign and equal magnitude, $\Omega_z(X)=-\Omega_z(Y)$ (Fig.~{\color{blue}S5}~\cite{SM}).
Under hole doping, carriers from the two valleys acquire opposite transverse velocities; owing to valley-spin locking, this transverse response constitutes an intrinsic valley-spin Hall effect~\cite{SHE1,SHE2,SHE3,SHE4}.

However, the two valley contributions cancel in the charge channel, resulting in a vanishing AHC. We evaluate the AHC as~\cite{Berry1,Berry2}
$$ 
\sigma_{xy}=-\frac{e^2}{\hbar} \sum_{n} \int_{\text{BZ}} \frac{d^2 k}{(2\pi)^2} f(k)\Omega(k),
$$
where $f(k)$ is the Fermi-Dirac distribution function. This cancellation leaves the zero-field system in a pure valley-spin Hall state with no net charge Hall current.

A finite gate field breaks the symmetry that enforces this cancellation. For $\mathcal{E}_z>0$, the gate-induced valley splitting makes one valley dominate the hole-doped transport window, producing an asymmetric Berry-curvature contribution and a positive $\sigma_{xy}$ near the valence-band edge. Reversing the gate polarity reverses the valley ordering and the Berry-curvature imbalance, and $\sigma_{xy}$ changes sign [Fig.~\ref{fig:three}{\color{blue}(a)}]. Thus, the gate field converts the zero-field pure valley-spin Hall response into a net charge Hall response. The sign of the EHE is controlled by the gate polarity, while the spin character of the conducting holes is selected indirectly through spin-valley locking.

This transport switching demonstrates the central advantage of our mechanism. The system remains magnetically compensated, yet the gate field selects one layer-polarized valley to dominate charge Hall transport. Consequently, the valley index, the dominant spin character, and the sign of the Hall voltage can all be controlled electrically. The SVLC-enabled EHE thus bridges the low-stray-field advantage of compensated magnets with the electrical readout characteristic of $\mathcal{T}$-symmetry-broken valley systems.


\paragraph{{\color{blue} Gate-switchable magneto-optical Kerr effect—}}%
Beyond electrical transport, SVLC enables an electrically switchable magneto-optical Kerr effect (MOKE). At $\mathcal{E}_z=0$, the symmetry-related $X$ and $Y$ valley states yield canceling optical Hall responses, so $\sigma_{xy}(\omega)$ vanishes. A finite gate field lifts the valley degeneracy and removes this cancellation, thereby allowing a nonzero $\sigma_{xy}(\omega)$ and a gate-reversible Kerr response.

For a 2D system on a nonmagnetic substrate, the complex Kerr angle is given by~\cite{kerr-gs1,kerr-gs2,kerr-zxd1,kerr-ds}
$$
\phi_K=\theta_K+i\epsilon_K=i \frac{2\omega d}{c} \frac{\sigma_{xy}}{\sigma_{xx}^s},
$$
where $\theta_K$ and $\epsilon_K$ are the Kerr rotation and ellipticity, respectively, and $d$ is the effective thickness of the magnetic monolayer. We take SiO$_2$ as the substrate, with $\sigma_{xx}^s=i(1-n^2)\omega/4\pi$ and refractive index $n=1.546$. The quantity $\sigma_{xy}$ denotes the off-diagonal component of the optical conductivity tensor.

The symmetry change is reflected in the optical conductivity tensor (OCT). With a finite gate field, the fFIM allows an antisymmetric off-diagonal component,
$$
\text{OCT}\left(\mathcal{E}_z \neq 0\right)=
\begin{pmatrix}
\sigma_{xx} & \sigma_{xy} & 0 \\
-\sigma_{xy} & \sigma_{xx} & 0 \\
0 & 0 & \sigma_{zz}  
\end{pmatrix},
$$
whereas the zero-field altermagnetic state forbids this component,
$$
\text{OCT}\left(\mathcal{E}_z =0\right)=
\begin{pmatrix}
\sigma_{xx} & 0 & 0 \\
0 & \sigma_{xx} & 0 \\
0 & 0 & \sigma_{zz}
\end{pmatrix}.
$$

The calculated spectra confirm this symmetry analysis (Fig.~\ref{fig:three}{\color{blue}(b)} and Fig.~{\color{blue} S6}~\cite{SM}). At $\mathcal{E}_z=0$, the off-diagonal optical conductivity and the Kerr response vanish. A finite gate field activates both Kerr rotation and ellipticity, and reversing the gate polarity reverses their signs. The maximum Kerr rotation exceeds 1 mrad, comparable to that reported for monolayer CrI$_3$~\cite{kerr-CrI3-1mrad} and well above typical experimental detection limits. The calculated MOKE response therefore demonstrates electric-field switching of the Kerr rotation angle, whose sign reverses with the gate polarity.

\paragraph{{\color{blue}Conclusion—}}%

In summary, we have uncovered a hidden spin-valley Hall response in altermagnets and demonstrated that it can be electrically exposed and switched. Although the net charge Hall response vanishes by symmetry, the opposite-spin valleys carry compensating Hall currents of opposite sign. SVLC links these valleys to opposite layer polarizations, so that a perpendicular gate field removes this compensation and converts the hidden response into a gate-switchable EHE. Using cSLG theory, we established the symmetry criteria for this mechanism, screened all 158 altermagnetic cSLGs, identified 32 that support SVLC, and predicted monolayer Mg$_2$Os$_2$(SO$_5$)$_2$ and the Ca(CoN)$_2$ family as candidate materials. First-principles calculations for Mg$_2$Os$_2$(SO$_5$)$_2$ confirm reversible valley splitting, a gate-switchable Hall response, and a Kerr rotation exceeding 1 mrad that reverses sign together with the electric Hall signal. These results establish altermagnets with SVLC as a concrete platform for all-electrical valley-spin manipulation, gate-switchable Hall transport, and electrically reversible magneto-optics. More generally, they demonstrate how hidden responses in symmetry-compensated magnets can be exposed by coupling spin, valley, and layer degrees of freedom.

\begin{acknowledgments}
\paragraph{Acknowledgments—}%
This work is supported by the National Key R\&D Program of China (Grants No.2025YFA1411200, No. 2024YFA1409002), the National Natural Science Foundation of China (Grants No. 12304188, No. 12234003). The Natural Science Foundation of Beijing (Grant No. 1252029), the Beijing National Laboratory for Condensed Matter Physics (Grant No. 2024BNLCMPKF006), and the Beijing Institute of Technology Research Fund Program for Young Scholars.
This work was also funded by China Postdoctoral Science Foundation (Certificate No. 2026M794613).
\end{acknowledgments}

\bibliography{ref}



\end{document}